\documentclass[journal]{rmaa}

\title{Radio Continuum Sources associated with \\
the HH~92 and HH~34 Jets}

 \author{Luis F. Rodr\'\i guez\altaffilmark{1,2}, Bo Reipurth\altaffilmark{3,4}, and Hsin-Fang Chiang\altaffilmark{3,4}}
  \altaffiltext{1}{Centro de Radioastronom\'{\i}a y Astrof\'{\i}sica, Universidad
Nacional Aut\'onoma de M\'exico, Campus Morelia} 

\altaffiltext{2}{Astronomy Department, Faculty of Science, King Abdulaziz University, P.O.
Box 80203, Jeddah 21589, Saudi Arabia}

\altaffiltext{3}{Institute for Astronomy, University of Hawaii at Manoa, Hilo, HI, USA}

\altaffiltext{4}{NASA Astrobiology Institute, University of Hawaii at Manoa, Hilo, HI, USA} 

\fulladdresses{
\item Luis F. Rodr\'\i guez: Centro de Radiostronom\'{\i}a
y Astrof\'\i sica, 
UNAM, A. P. 3-72, 
(Xangari), 58089 Morelia, Michoac\'an, M\'exico
(l.rodriguez@crya.unam.mx).

\item Bo Reipurth: Institute for Astronomy,
University of Hawaii at Manoa, 640 North Aohoku Place, Hilo, HI 96720, USA
 (reipurth@ifa.hawaii.edu).

\item Hsin-Fang Chiang: Institute for Astronomy,
University of Hawaii at Manoa, 640 North Aohoku Place, Hilo, HI 96720, USA
 (hchiang@ifa.hawaii.edu).
}

\shortauthor{Rodr\'\i guez, Reipurth, Chiang}
\shorttitle{Radio Sources at HH~92 and HH~34}

\SetVolume{1} \SetFirstPage{001} \SetYear{2014}
\ReceivedDate{2014 April 14} 
\AcceptedDate{2014 May 23} 

\resumen{Presentamos observaciones de alta resoluci\'on angular y alta
  sensitividad en el radiocontinuo a 8.46 GHz (3.6 cm) hacia el n\'ucleo del
  flujo HH~92 hechas con el Very Large Array en 2002-2003 y con el
  Expanded Very Large Array en 2011.  Detectamos un grupo de tres
  fuentes compactas distribuidas en una regi\'on con 2$''$ de
  extensi\'on y discutimos su naturaleza. Concluimos que una de las fuentes (VLA 1)
es la excitadora del flujo gigante asociado con HH~92. En el caso de
HH~34 presentamos nuevas observaciones a 43.3 GHz (7 mm) que revelan la presencia
de una estructura asociada con la fuente excitadora y alargada perpendicularmente
al chorro \'optico altamente colimado en la regi\'on. Proponemos que la
fuente de 7 mm es un disco circunestelar con
radio de $\sim$80 AU y masa de $\sim$0.21 $M_\odot$.}
 
\abstract{We present high angular resolution, high sensitivity 8.46
  GHz (3.6 cm) radio continuum observations made toward the core of the HH~92
  outflow with the Very Large Array in 2002-2003 and with the Expanded
  Very Large Array in 2011.  We detect a group of three compact
  sources distributed in a region 2$''$ in extension and discuss
  their nature. We conclude that one of the objects (VLA 1) is the exciting source of the
giant outflow associated with HH~92. In the case of HH~34 we present new 43.3 GHz (7 mm)
observations that reveal the presence of a structure associated with the
exciting source and elongated perpendicular to the highly collimated optical jet in the region.
We propose that this 7 mm source is a circumstellar disk with radius of $\sim$80 AU
and mass of $\sim$0.21 $M_\odot$.}
      

\begin{document}

\maketitle

\section{Introduction}

It is well-established that the typical outcome of the collapse of a cloud is a small multiple system
(e.g. Duch\^ene
et al. 2006; Reipurth et al. 2014).  In particular, at centimeter
wavelengths the studies of Reipurth et al.  (2002a; 2004) have shown
that the cores of regions associated with outflow activity typically
reveal the presence of several radio continuum sources, each one
usually believed to be tracing the presence of a recently formed star.
However, these compact radio sources could also trace collisionally
ionized regions, that is, the radio equivalent of Herbig-Haro objects,
sources without an embedded star.

In the case of forming stars of low and intermediate mass, the
observed radio emission directly associated with the young star has
two possible origins. A first class of sources are those with resolved
structure at the scale of tenths of arc seconds that are usually
emitting free-free (thermal) radiation from ionized, collimated
outflows (e.g., Curiel et al. 1993). A second class of sources are young stars with
magnetospheric activity that produces detectable gyrosynchrotron
(non-thermal) emission (e.g., Dzib et al. 2013). The determination
of the characteristics of these sources can be of great help in
improving our understanding of the region studied, locating precisely
the position of young stars with either outflow or magnetospheric
activity.  In particular, the detection of elongated free-free
sources, sometimes called thermal jets (Anglada 1996; Rodr\'\i guez
1997; Rodr\'\i guez et al. 1999; Rodr\'\i guez et al. 2000), locates the exciting source and the
origin and orientation of the outflow at sub-arcsec scales.
At the shortest wavelengths, e.g. 7 mm, dust emission also becomes
detectable, allowing additional high resolution imaging of
circumstellar material (Rodr\'\i guez et al. 2008a).

\subsection{The HH 92 Jet and its Driving Source}

\begin{figure}
\centering
\includegraphics[scale=0.34, angle=0]{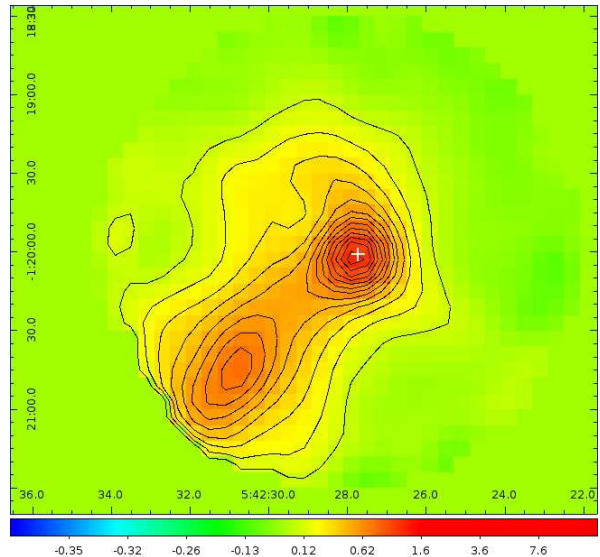}
\caption{An 850~$\mu$m SCUBA map of the filamentary cloud clump LBS~30
  in which the HH 92 driving source is located (white cross). The
  elongated cloud clump has fragmented into two separate cores, of
  which only the northwestern has begun to form stars (Miettinen \&
  Offner 2013). The cross has the same dimensions (4.5$"$ $\times$
  4.5$"$) as the panels in Figure~\ref{fighh92vla}.  }
 \label{scubafig}
\end{figure}

Bally, Reipurth, \& Aspin (2002) reported the detection of a highly
collimated, low-excitation jet, HH~92, that emerges (with a position
angle of $311^\circ$) from an asymmetric infrared reflection nebula
spatially related with the source IRAS~05399-0121, an embedded
low-mass class~I protostar with a bolometric luminosity of 10
$L_\odot$.  In their study, Bally et al. (2002) propose that HH~92,
together with the well-known objects HH 90/91 and HH~93 (Reipurth
1985; 1989) trace a single giant outflow lobe $\simeq$3 pc in length.
The driving source is located in a cloud clump labeled as LBS~30 (Lada
et al. 1991) with two cores, as seen in an 850~$\mu$m SCUBA image
(Figure~\ref{scubafig}) from the SCUBA Legacy Survey (Di Francesco et
al. 2008).  Similar structure is seen in the 350~$\mu$m maps of
Miettinen \& Offner (2013).

However, despite several early efforts the exciting source of this
giant outflow was initially not firmly identified (Gredel, Reipurth,
\& Heathcote 1992; Davis, Mundt, \& Eisl\"oeffel 1994; Reipurth et al.
1993). With the advent of the Spitzer and Herschel missions, an
accurate position and photometry were achieved (Megeath et al. 2012;
Manoj et al. 2013). Most recently, the source has been studied in
detail at sub-mm wavelengths by Miettinen \& Offner (2013), who give
further references to the literature.

In this paper we present sensitive, high angular resolution Very Large
Array and Expanded Very Large Array observations at 8.46~GHz (3.6 cm)  of the
core of the region that reveal the presence of a compact group of
radio sources.  A distance of 415 pc to the L1630 cloud, where HH~92
is embedded, is adopted (Anthony-Twarog 1982).

\subsection{HH 34 and its Driving Source}

The Herbig-Haro object HH~34 was first noted by Guillermo Haro, see
Herbig (1974). It is part of a major complex of shocks, including a
highly collimated jet (Reipurth et al. 1986), and distant bow shocks
that make the complex span across almost 3~pc (Devine et al. 1997).
Proper motions and radial velocities indicate a dynamic age of the
entire complex of about 10,000~yr (Devine et al. 1997, Reipurth et al.
2002b). The driving source is not optically visible, although a
compact reflection nebula is visible at the base of the jet, and
spectroscopy of this nebula shows that the source is a rich
emission-line T~Tauri star (Reipurth et al. 1986). Imaging with NICMOS
on HST reveals the driving source in the H- and K-bands (Reipurth et
al. 2000). The source was also detected with low spatial resolution in
the radio continuum at 3.6~cm (Rodr\'\i guez \& Reipurth 1996).  HH~34
is located in the northern part of the L1641 cloud, just south of the
Orion Nebula cluster, whose distance has been determined as 414$\pm$7
pc by Menten et al. (2007) and 418$\pm$6 pc by Kim et al. (2008), see
also the discussion in Muench et al. (2008). We here use the mean
distance of 416~pc.

In this paper we present a deep high angular resolution radio
continuum VLA map obtained at 7~mm.

\begin{figure*}
\centering
\includegraphics[scale=0.60, angle=0]{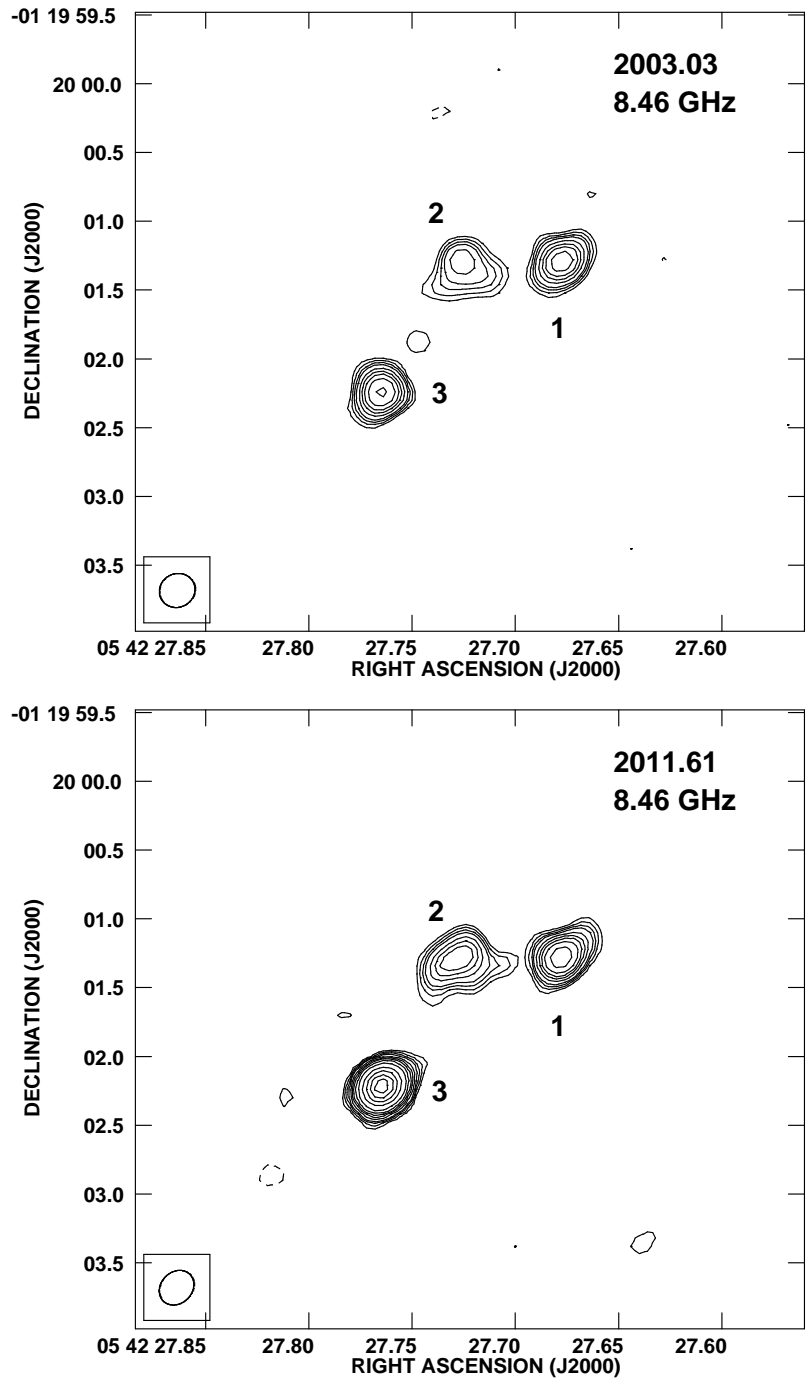}
\caption{Contour images of the 8.46 GHz (3.6 cm)  continuum emission from the
  core of the HH~92 outflow. (Top) Image for the 2003.03 epoch.  The
  contours are -4, -3, 3, 4, 5, 6, 8, 10, 12, 15, and 20 times
  12.4 $\mu$Jy beam$^{-1}$, the rms noise of the image.  The half power contour of
  the synthesized beam ($0\rlap.{''}26 \times 0\rlap.{''}24$ with a
  position angle of $-62^\circ$) is shown in the bottom left corner.
  (Bottom) Image for the 2011.61 epoch.  The contours are -4, -3, 3,
  4, 5, 6, 8, 10, 12, 15, 20, 30, 40, 50, and 60 times 8.9 $\mu$Jy beam$^{-1}$, the
  rms noise of the image.  The half power contour of the synthesized
  beam ($0\rlap.{''}27 \times 0\rlap.{''}23$ with a position angle of
  $-45^\circ$) is shown in the bottom left corner.  }
  \label{fighh92vla}
\end{figure*}

\section{Observations}

\subsection{HH 92}

The Very Large Array (VLA) observations were made at 8.46 GHz (3.6 cm)  in the B
(during epoch 2002 August 1) and A (during epoch 2003 June 21)
configurations.  The VLA is part of the NRAO\footnote{The National
  Radio Astronomy Observatory is operated by Associated Universities
  Inc. under cooperative agreement with the National Science
  Foundation.}.  In both epochs the source J1331+305 was used as an
absolute amplitude calibrator (with an adopted flux density of 5.21
Jy) and the source J0541-056 was used as the phase calibrator (with
bootstrapped flux densities of 0.879$\pm$0.003 and 0.958$\pm$0.007 Jy
for the first and second epochs, respectively).  The data were reduced
separately for each epoch using the standard VLA procedures and later
concatenated to produce images of high sensitivity and high angular
resolution. We refer to these concatenated data as having epoch
2003.03, the average epoch of the two observations.

With the purpose of studying the time evolution and kinematics of
these sources, we made Expanded Very Large Array (EVLA) observations
of the same region during 2011 August 11 in the A configuration.  We
refer to these data as having epoch 2011.61, 8.58 years after the
first observations.  The EVLA observations were made at 8.46 GHz (3.6 cm) using
two intermediate frequency (IF) bandwidths of 128 MHz each, separated
by 128 MHz, and containing both circular polarizations. Each IF was
split in 64 channels of 2 MHz each. For the continuum images we
averaged the central 54 channels, rejecting five channels at each end
of the bandwidth.  The data reduction was made using the software
package AIPS of NRAO, following the recommendations for EVLA data
given in Appendix E of its CookBook (that can be found at 
http://www.aips.nrao.edu/cook.html).  The source J0137+3309 was used
as an absolute amplitude calibrator (with an adopted flux density of
3.15 Jy) and the source J0541-056 was used as the phase calibrator
(with a bootstrapped flux density of 0.699$\pm$0.003).

In Figure~\ref{fighh92vla} we show the images of the HH~92 region at
the two epochs studied.  The top one is made from the 2003.03 data with
the \sl (u,v) \rm weighting parameter ROBUST (Briggs 1995) of the task IMAGR
set to -5.  The bottom one is made from the 2011.61 data with the
\sl (u,v) \rm weighting parameter ROBUST of the task IMAGR set to 0.  The
different weightings were needed to produce images of very similar
angular resolution, given that the 2003.03 data contains both A and B
configuration observations, while the 2011.61 data contains only A
configuration observations.  These images clearly show three sources
that we label as sources VLA 1, 2, and 3.  The three sources are
distributed in a region of $\sim 2{''}$ in extent (830 AU at a
distance of 415 pc).  In Table 1 we list the parameters of these three
sources, taken from the images at both epochs.
 
\subsection{HH 34}

To obtain an image of the core of HH~34 at 43.3 GHz (7 mm) we used
two data sets from the VLA archive. The first data set was taken on 2004 October 05
in the A configuration under project AR552 and the second data set was taken on
2006 August 29 in the B configuration under project AK634. Both observations were
made with an effective bandwidth of 100 MHz. For the first epoch 0713+438 was used as flux calibrator
(with an adopted flux density of 0.29 Jy) and 0541-056 as phase calibrator, with a bootstrapped 
flux density of 0.61$\pm$0.01 Jy. For the second epoch 1331+305 was used as flux calibrator 
(with an adopted flux density of 1.45 Jy) and 0541-056 as phase calibrator, with a bootstrapped
flux density of 1.88$\pm$0.06 Jy. The data were calibrated following the standard VLA
procedures for high frequency observations and the two epochs were concatenated to obtain
an image with better sensitivity. In Figure 3 we show 
the resulting image, made with a ROBUST weight of 5 (Briggs 1995) to emphasize sensitivity.
An extended source is clearly detected, with peak position $RA(2000) = 05^h~ 35^m~ 29\rlap.{^s}846 \pm 0\rlap.{^s}001$,
$DEC(2000) = -06^\circ~ 26'~ 58\rlap.{''}08 \pm 0\rlap.{''}02$, a total flux density of 3.5$\pm$0.4 mJy and 
deconvolved dimensions
of $0\rlap.{''}39 \pm 0\rlap.{''}05 \times 0\rlap.{''}25 \pm 0\rlap.{''}04; PA = 55^\circ \pm 13^\circ$.

\begin{table*}[htbp]
\small
  \setlength{\tabnotewidth}{1.8\columnwidth} 
  \tablecols{6} 
  \caption{Radio Sources at the Core of the HH~92 Outflow$^{\lowercase{a}}$}
  \begin{center}
    \begin{tabular}{lcccccc}\hline\hline
& &\multicolumn{2}{c}{Position$^b$} & Total Flux 
& & \\
\cline{3-4} 
VLA & Epoch &  $\alpha$(J2000) & $\delta$(J2000) & Density (mJy) & 
Deconvolved Angular Size$^c$ \\ 
\hline
1 & 2003.03 & 05 42 27.678 & -01 20 01.29 & 0.31$\pm$0.02 
& $0\rlap.{''}23 \pm 0\rlap.{''}03 \times 0\rlap.{''}08 \pm 0\rlap.{''}03;~ +130^\circ\pm 9^\circ$  \\
1 & 2011.61 & 05 42 27.677 & -01 20 01.27 & 0.29$\pm$0.02 
& $0\rlap.{''}20 \pm 0\rlap.{''}03 \times 0\rlap.{''}08 \pm 0\rlap.{''}03;~ +122^\circ\pm 10^\circ$  \\
 & & & & & \\
2 & 2003.03 & 05 42 27.724 & -01 20 01.35 & 0.28$\pm$0.03
& $0\rlap.{''}36 \pm 0\rlap.{''}05 \times 0\rlap.{''}26 \pm 0\rlap.{''}06;~ +106^\circ\pm 34^\circ$  \\
2 & 2011.61 & 05 42 27.728 & -01 20 01.31 & 0.28$\pm$0.02
& $0\rlap.{''}37 \pm 0\rlap.{''}04 \times 0\rlap.{''}34 \pm 0\rlap.{''}04;~ +106^\circ\pm 12^\circ$  \\
 & & & & & \\
3 & 2003.03 & 05 42 27.765 & -01 20 02.24 & 0.34$\pm$0.02
& $0\rlap.{''}17 \pm 0\rlap.{''}04 \times 0\rlap.{''}10 \pm 0\rlap.{''}04;~ +162^\circ\pm 28^\circ$  \\
3 & 2011.61 & 05 42 27.765 & -01 20 02.22 & 0.64$\pm$0.02
& $0\rlap.{''}06 \pm 0\rlap.{''}02 \times 0\rlap.{''}04 \pm 0\rlap.{''}03;~ +96^\circ\pm 45^\circ$  \\

\hline\hline
\tabnotetext{a}{Parameters taken from the images at each epoch.}
\tabnotetext{b}{Units of right 
ascension are hours, minutes, and seconds
and units of declination are degrees, arcminutes, and arcseconds. Positional accuracy
is estimated to be $0\rlap.{''}01$.}
\tabnotetext{c}{Major axis $\times$ minor axis; position angle.}
    \label{tab:1}
    \end{tabular}
  \end{center}
\end{table*}

\section{HH 92: ~~Results and Interpretation}

\subsection{Proper Motions}

The positions of the three sources at the two epochs (see Table 1) are
approximately coincident within the positional error (1$\sigma \simeq
0\rlap.{''}01$). Using the distance of 415 pc and the time separation
of 8.58 years between observations, we set a 3-$\sigma$ upper limit of
$\leq$7~km s$^{-1}$ for the proper motions of the sources. This rules
out an interpretation in terms of some of them being classic HH knots,
since such objects exhibit proper motions an order of magnitude or
more larger. Most probably, the radio sources are tracing young stars
or stationary HH knots.

\subsection{Flux Density Variations as a Function of Time}

The sources VLA 1 and VLA 2 do not exhibit flux density variations
within the observational noise of $\sim$0.02-0.03 mJy. This sets an
upper limit of $\sim$10\% to any possible flux density variations.  In
contrast, source VLA 3 shows a strong increase of a factor of 2, from
0.34 to 0.64 mJy. Furthermore, the deconvolved size of this source
decreased for the second epoch, suggesting that most of the flux
density increase took place in a very compact region, perhaps due to
gyrosynchrotron activity in VLA 3.

\subsection{Morphology of the Sources}

The deconvolved angular dimensions of the sources VLA 2 and VLA 3 are
consistent, within noise, with a spherical geometry. In contrast, VLA 1 is clearly
elongated, with its major axis $\sim$3 times larger than its minor
axis. The average position angle of this source (from the two epochs
of observation) is $126^\circ \pm 7^\circ$, coincident within modulo
$180^\circ$ with the position angle of the optical jet, $311^\circ$
(Bally et al. 2002).

\subsection{Interpretation}

The resolution of the HH 92 driving source into
three radio continuum sources can be interpreted in
terms of the sources being either young stars or 
representing thermal emission from HH shocks.
The radio emission could also have a non-thermal origin,
as has been observed in the knots of some jets (e.g. Carrasco-Gonz\'alez et al. 2010), 
but we do not have the multifrequency data
required to test this possibility.

Recent submm observations using the SMA interferometer by Chiang et
al. (in prep.) have revealed that there is only one 850~$\mu$m source
in the field of the triple system, namely VLA~1. Neither VLA~2 nor
VLA~3 are detected. This strongly suggests that VLA~1 is the driving
source, and is consistent with the elongation along the flow axis seen
for VLA~1 (see above).

So what are VLA~2 and 3? Both lie approximately within the redshifted
outflow lobe from VLA~1, which would indicate that they are HH knots.
However, our two-epoch observations spanning 8~1/2 yr reveal no
measurable proper motions at all, and put very stringent limits on the
proper motions of less than 7~km~s$^{-1}$. Since HH flows typically
have space velocities of 100 km~s$^{-1}$ or more, such slow motions
would indicate that the motion is mostly along the line of sight.
However, this is inconsistent with the giant flow dimensions of about
3~pc, which points towards an outflow axis lying close to the plane of
the sky. It follows that if VLA~2 and 3 are shocks, then they are
stationary shocks, such as would occur if small ambient cloudlets are
interacting with a strong stellar wind or outflow (Schwartz 1978).
Possible stationary shocks with no detectable proper motions along the flow have been
found in regions of high ambient density (e.g., Mart\'\i\ et al. 1995;
Rodr\'\i guez et al. 2008b).
This interpretation gains weight because at least VLA~3 shows strong
flux-variability over the time span of the observations.

Alternatively, a possible scenario would be that the HH 92 driving source forms a small 
non-hierarchical triple stellar system, and that an outflow from the active source VLA 1 
runs over the two nearby stellar components of the triple system, creating shocks where 
the outflow rams into the circumstellar material of these two stars.  A similar case has 
recently been identified in the young IRAS 16293–2422 triple system (Girart et al. 2014). 
This scenario is not inconsistent with the lack of detection of VLA 2 and 3 at submm wavelengths, 
because not all very young stars are detectable at sub-mm wavelengths, since the cool 
circumstellar envelopes merely re-emit instantaneously the  radiation from the central 
protostars, and if the accretion is dormant, the re-emission from the envelopes will be weak. 

The presently available observations do not allow us to distinguish
between the two scenarios in which VLA~2 and 3 are either small
cloudlets or protostellar companions.

\section{HH 34: ~~Results and Interpretation}

\begin{figure}
\centering
\includegraphics[scale=0.43, angle=0]{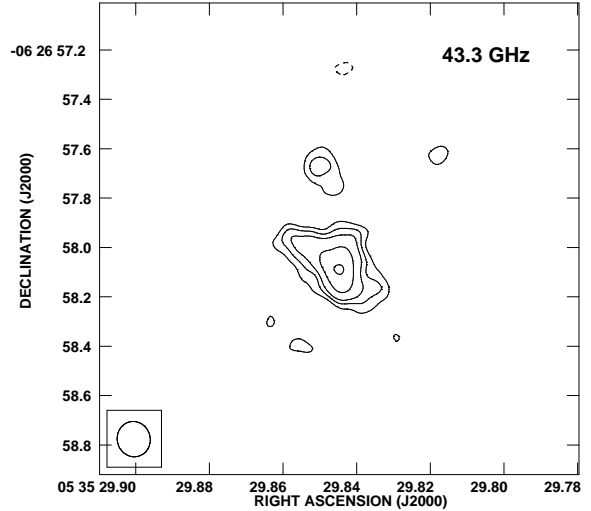}
\caption{A 43.3 GHz (7~mm) radio continuum map of a 1.9$"$ $\times$ 1.9$"$ area
  around the HH~34 driving source. The elongation of the source is
  perpendicular to the jet axis as seen in Figure~\ref{hh34hstfig} and
  is likely to represent cool dust in a circumstellar disk. At the
  assumed distance of 416~pc, the 0.4$"$ width of the disk corresponds
  to about 160~AU. The
  contours are -4, -3, 3, 4, 5, 7, and 9 times
  77 $\mu$Jy beam$^{-1}$, the rms noise of the image.  The half power contour of
  the synthesized beam ($0\rlap.{''}14 \times 0\rlap.{''}13$ with a
  position angle of $14^\circ$) is shown in the bottom left corner. }
 \label{hh34fig}
\end{figure}

\begin{figure}
\includegraphics[scale=0.27, angle=0]{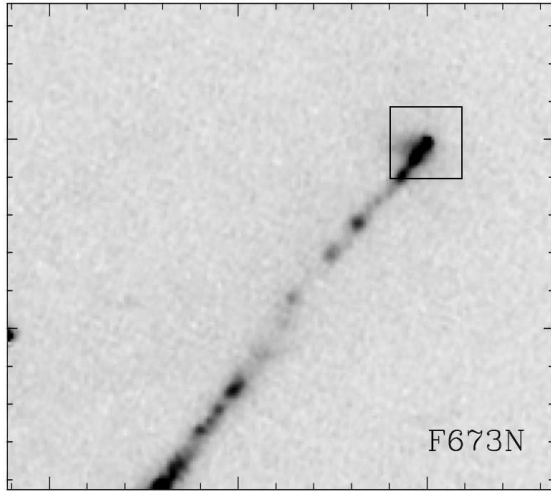}
\caption{A WFPC2 HST image of the base of the HH 34 jet obtained
  through a narrow-band filter transmitting the [SII] 6717/6731
  doublet lines. Tick marks are in steps of one arcsec, and the field
  of view is about 12$"$ $\times$ 14$"$. The box outlines the 1.9$"$
  $\times$ 1.9$"$ field shown in 7~mm radio continuum in
  Figure~\ref{hh34fig}. Image from Reipurth et al. (2002b). }
 \label{hh34hstfig}
\end{figure}

Our 7~mm map of the HH~34 driving source is seen in
Figure~\ref{hh34fig}. When compared to the beam, the source is clearly
extended in a NE-SW direction, but there is also an extension almost
perpendicular to the main elongation. In Figure~\ref{hh34hstfig} we
show the area of our 7~mm map shown in Figure~\ref{hh34fig} as a
square superposed on a high-resolution optical image ([SII] 6717/6731
transitions) of the HH~34 jet. It is immediately evident that the main
elongation of the source is perpendicular to the jet axis, suggesting
that what we see is dust emission from a circumstellar disk. The disk
has a total width of 0.4~arcsec, which corresponds to a radius of
$\sim$80~AU, matching well with expectations for Class~I disks.  At
7~mm, both dust emission and free-free emission can be detected
(Rodr\'\i guez et al. 2008a), and so it
seems likely that the elongation we see perpendicular to the disk axis
results from the jet, which is monopolar at optical wavelengths, but
bipolar at mid-infrared wavelengths (Raga et al. 2011). 

We estimate the total mass of the 7 mm source
by assuming optically thin, isothermal dust emission and a gas-to-dust ratio of 100
(Sodroski et al. 1997). The total mass opacity coefficient at 7 mm is poorly known and
values between $8 \times 10^{-3}$ cm$^2$ g$^{-1}$ (Rodr\'\i guez et al. 2007) and
$2 \times 10^{-3}$ cm$^2$ g$^{-1}$ (Isella et al. 2014) can be found in the literature.
Assuming an average value of $5 \times 10^{-3}$ cm$^2$ g$^{-1}$, the  
total mass can be roughly estimated to be:

\footnotesize
$$ \Biggl[{{M} \over {M_\odot}}\Biggr] = 0.17 
\Biggl[{{S_\nu} \over {mJy}}\Biggr]
\Biggl[{{T} \over {100~K}}\Biggr]^{-1}
\Biggl[{{D} \over {kpc}}\Biggr]^{2},$$

\normalsize

\noindent where $S_\nu$ is the flux density at 7 mm,
$T$ is the dust temperature and $D$ is the distance 
to the source. Assuming
$T$ = 50 K and a distance of 416 pc, we estimate a mass
of $\sim0.21~M_\odot$ for the HH 34 disk.  

Almost precisely $\sim0\rlap.{''}4$ north of the HH 34 source there is a weak signal,
which is possibly real since it is detected at a 4-$\sigma$ level, but which will require confirmation from
additional observations. If real, it is located 180~AU in projection
from the main source. Reipurth (2000) has suggested that all giant HH
flows are the result of dynamical interactions in and disintegration
of triple or small multiple systems. In this interpretation, the HH~34
source should be a close binary with separation of order 10~AU. No
existing data of the HH~34 source has sufficient spatial resolution to
detect such a close binary. In addition to the companion possibly
detected here, another very faint companion (``HH~34~X'') was detected
in the K-band with HST by Reipurth et al. (2000), about 5$"$ west of
the jet source. Finally, a faint variable star (``Star~B'') is located
11$"$ SE of the driving source and seen at the left edge of Fig. 4 (Reipurth et al. 1986, 2002b).
Altogether, the HH~34 driving source may have several more distant
companions.

\section{Conclusions}

We have presented VLA and EVLA observations made at 8.46 GHz (3.6 cm)  of the
core of the HH~92 outflow, made at two epochs separated by 8.58 years.
We detect a group of three compact radio sources, of which VLA~1
drives the HH~92 jet. The other two sources, VLA~2 and 3, may be
stationary shocks surrounding either small cloudlets or protostars. 

Additional VLA observations obtained at 7~mm of the driving source of
the HH~34 jet have revealed a circumstellar disk perpendicular to the
jet axis and with a radius of $\sim$80~AU and mass of $\sim$0.21 $M_\odot$.


\acknowledgments LFR acknowledges the support of DGAPA, UNAM, and of
CONACyT (M\'exico).  BR and HFC acknowledge support by the National
Aeronautics and Space Administration through the NASA Astrobiology
Institute under Cooperative Agreement No. NNA09DA77A issued through
the Office of Space Science.  This research has made use of the SIMBAD
database, operated at CDS, Strasbourg, France.


\end{document}